\documentstyle[sprocl,psfig]{article}

\def\be{\begin{equation}}
\def\ee{\end{equation}}
\def\bea{\begin{eqnarray}}
\def\eea{\end{eqnarray}}

\begin{document}


\title{\bf DYSON-SCHWINGER EQUATION APPROACH TO THE QCD\\ 
DECONFINEMENT TRANSITION AND J/$\psi$ DISSOCIATION}

\author{D. B. BLASCHKE, G. R. G. BURAU}

\address{Fachbereich Physik, Universit\"at Rostock, 
D--18051 Rostock, Germany} 

\author{M. A. IVANOV} 
\address{Bogoliubov Laboratory for Theoretical Physics,
Joint Institute for Nuclear Research\\
14 19 80 Dubna, Russian Federation}

\author{Yu. L. KALINOVSKY}

\address{
Laboratory of Computing Techniques and Automation\\
Joint Institute for Nuclear Research, 
14 19 80 Dubna, Russian Federation}

\author{P. C. TANDY}

\address{Centre for Nuclear Research, Department of Physics, 
Kent State University,\\Kent OH 44242, USA}


\maketitle
\abstracts{We consider an extension of the finite-temperature 
Dyson-Schwinger equation (DSE) approach to heavy mesons and quarkonia 
and apply it to calculate the cross section for the J/$\psi$ breakup 
reaction J/$\psi~+~\pi~\longrightarrow~$ D~+~$\bar{\rm D}~$.
We study the effects of chiral symmetry restoration in the light quark 
sector on this process and obtain a critical enhancement of the reaction rate
at the chiral/ deconfinement transition. 
Implications for the kinetics of charmonium production in ultrarelativistic 
heavy-ion collisions are discussed with particular emphasis on the recently 
observed anomalous J/$\psi$ suppression as a possible signal for quark matter 
formation.}

\section{Introduction}
\begin{figure}[hbt]
{\psfig{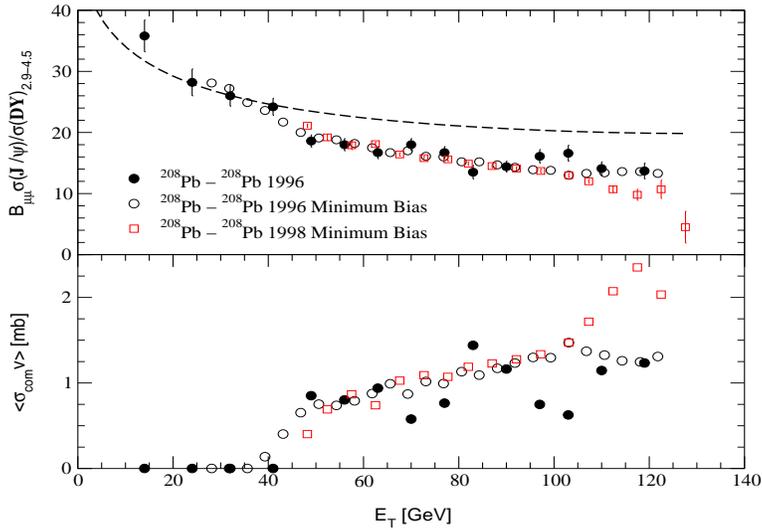}}
\caption{Upper panel: J/$\psi$ production in Pb-Pb collisions at $158$ AGeV 
\protect \cite{ramello,abreu98} compared to the extrapolation from p-A and S-U 
collisions using a Glauber model (dashed line) with nuclear J/$\psi$ 
absorption. Lower panel: Comover absorption cross sections extracted from the 
difference of observed data to the nuclear absorption extrapolation.}
\label{na50}
\end{figure}

Recent results of the NA50 collaboration at CERN SPS on anomalous
J/$\psi$ suppression in Pb-Pb collisions at 158 A GeV \cite{ramello,abreu98}
have renewed the quest for a proper description of charmonium
production in heavy-ion collisions.
It has been emphasized that a threshold effect in the
E$_T$ dependence of J/$\psi$ supression like the one observed  by NA50 
could be a signal for quark-gluon
plasma formation \cite{ms86,b91,bo96}, see Fig. \ref{na50}.
However, up to now there is no theory for the hadroproduction of charmonium 
and the present approaches have to be considered as parametrizations of the 
data. We provide here a reparametrization of this threshold effect where we
extract effective comover cross sections from the difference of the observed 
data points to the extrapolated nuclear absorption within a generalized 
Glauber model \cite{bbm98}, see Fig. \ref{na50}, lower panel. 
This comover cross section shows an onset at the transverse energy of 
$E_T\sim 40$ GeV which we interpret as a signal for a critical 
phenomenon in dense matter due to the chiral/deconfinement transition. 
Within the present paper, we want to investigate the charmonium dissociation 
cross section within a DSE model of QCD at high temperature and density 
\cite{anu98} and thus go beyond previous estimates of this quantity within 
a nonrelativistic potential model \cite{mbq}.
This is particularly important since pions as quasi Goldstone bosons 
are involved in the dissociation process and a nonrelativistic description of
these states fails to describe their wave functions accurately.
The DSE approach will be applied for the description of the temperature 
dependence of heavy meson masses, which can lead to a dramatic increase in the 
J/$\psi$ dissociation cross section, as has been demonstrated previously~ 
\cite{bbm98}.
This behaviour could contribute to an interpretation of the still puzzling 
experimental situation.

\section{Dyson-Schwinger equation approach to light and heavy mesons}

\begin{figure}[h]
\centerline{\psfig{figure=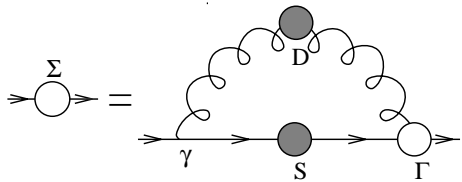,width=6cm}}
\caption{Self-energy diagram defining the fully dressed quark two-point 
function $S(p)$.}
\label{qdse}
\end{figure}

The nonperturbative features of low energy QCD such as dynamical chiral 
symmetry breaking and confinement can be successfully modeled by the 
truncation of the full DSE for the renormalized dressed quark propagator 
$S(p)$, see Fig. \ref{qdse}, to the rainbow level \cite{anu98,tandy}.
This approach has been systematically extended to the case of finite 
temperatures \cite{prl} and chemical potentials \cite{basti} within the 
imaginary time (Matsubara) formalism, where it gives insight into the 
interplay between chiral symmetry restoration and deconfinement found in 
Lattice QCD simulations \cite{karsch} and predicts the phase transition
parameters in the quark matter phase diagram \cite{thermo}.

Masses and decay properties of light mesons ($\pi,~\rho$) have been studied 
using the finite-temperature Bethe-Salpeter equations (BSEs) in the 
corresponding channels \cite{prl,rapid}. 
The DSE approach has also proven to be successful for the description of 
heavy meson observables \cite{ikmr} which play a crucial 
r\^{o}le in the present work.

The approach we present in the following will employ a confining,
separable model \cite{burden} recently extended to finite temperatures 
in the low-mass sector \cite{bkt,b+00}. 
In a rank-1 truncation the model gluon propagator reads
$D(p,q)=D_0~\varphi(p^2)\varphi(q^2)$,
and the renormalized quark propagator takes the form
\begin{eqnarray}
S(p) = -i \gamma_{\mu} p_{\mu} \sigma_V(p^2) + \sigma_S(p^2)
\end{eqnarray}
with $\sigma_{V}(p^2) = [p^2 + m^2(p^2)]^{-1},~ 
\sigma_{S}(p^2) = m(p^2)\sigma_{V}(p^2)$ 
and the dynamical quark mass function is given by
\begin{eqnarray}
m(p^2) &=& m_0 + \Delta m(T,\mu) \varphi (p^2).  
\end{eqnarray}
The solution of the quark DSE reduces to that of the gap equation
\begin{eqnarray}
\label{gap}
\Delta m(T,\mu)=\frac{16 D_0}{3}\int \frac{d^4q}{(2\pi)^4}
\varphi(q^2)\sigma_{S}(q^2)~,
\end{eqnarray}
where a nonvanishing mass gap $\Delta m(T,\mu)$ in the chiral limit  
$m_0=0$ signals spontaneous chiral symmetry breaking 
in the $T,\mu$ -plane of the phase diagram.
We choose a Gaussian formfactor $\varphi(p^2)=\exp(-p^2/\Lambda^2)$
and obtain a fit of $\pi$- and $\rho$- meson properties with
$D_0 = 227.8$~GeV$^{-2}$, $\Lambda = 0.60$~GeV, $m_0 = 7.2$~MeV.
The temperature dependence of the solution of the mass gap equation (\ref{gap})
results in a chiral restoration transition temperature $T_c=150$ MeV, see
upper panel of Fig. \ref{sigma}.
The model is confining for all $\Delta m(T,\mu)>\Lambda/\sqrt{2 {\rm e}}$ 
since it has then no solution of $p^2+m^2(p^2)=0$ for real $p^2$ 
(no quasiparticle poles).
For the heavy quark propagators the limit $m(p^2) = m_Q = {\rm const}$ 
will be used where $m_Q$ is the current mass of the heavy quark
\cite{ikmr}.

The BSE in ladder approximation for the mesonic bound states has the form
\begin{eqnarray}
\lambda(P^2) \Gamma_M (q,P) = - \frac{4}{3}
\int \frac{d^4p}{(2\pi)^4} D(q,p) \gamma_\mu S_{q_1}(p_+) \Gamma_M (p,P) 
S_{q_2}(p_-) \gamma_\mu
\end{eqnarray}
with $p_+ = p + \eta P,~ p_-= p - (1-\eta) P$, where
$\eta = 1/2$ for $m_{q_1} = m_{q_2}$ and $\eta = 1$ for $m_{q_1}\gg m_{q_2}$.
Here we will use the one-covariant meson vertex functions 
$\Gamma_M = \gamma_M N_M \varphi_M (p^2)$, 
where for the pseudoscalar mesons 
$\gamma_\pi=\gamma_D=\gamma_{\bar{D}}=i \gamma_5$ and for the vector mesons
$\gamma_\rho=\gamma_{D^*}=\gamma_{{\rm J}/\psi}=\gamma_\mu$ with the proper
normalization constants $N_M$. For the present study we will use the 
exploratory ansatz $\varphi_M (p^2)=\varphi(p^2)$.
The condition $\lambda (P^2 = -M_M^2) = 1$ 
leads to the meson mass formulae
\begin{eqnarray}
1 = \left. \frac{8}{3} D_0 \int \frac{d^4p}{(2\pi)^4} \varphi^2(p^2) 
\left[ V_M^{(V)}(p,P) \sigma_V^+ \sigma_V^- + V_M^{(S)}(p,P) \sigma_S^+ 
\sigma_S^- \right] \right|_{P=iM_M}~,
\end{eqnarray}
where the functions $V_M^{(V,S)}(p,P)$ can be found elsewhere \cite{b+00}.
With $m_Q = m_c = 1.844$~GeV the resulting masses for 
the heavy mesons (experimental value in brackets) are 
$M_D = 1.869$~GeV ~$((1.8693 \pm 0.0005)$~GeV), 
$M_{D^*} = 2.006$~GeV ~$((2.0067 \pm 0.0005)$~GeV) and 
$M_{J/\psi} = 3.459$~GeV ~$((3.09688 \pm 0.00004)$~GeV).

\section{Triangle diagram for the $D^* \rightarrow D \pi$ decay}

\begin{figure}[h]
\centerline{\psfig{figure=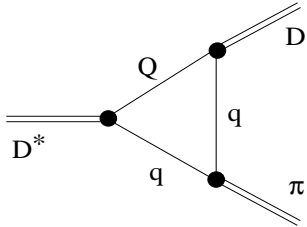,height=3cm,width=4cm}}
\caption{The quark loop diagram for $D^* \longrightarrow  D \pi$ decay vertex.}
\label{ddpi}
\end{figure}

We consider this vector-pseudoscalar-pseudoscalar decay as a 
test for our relativistic quark model. 
In the light meson sector this process describes the $\rho \pi \pi$ decay 
~\cite{tandy}, which is an important ingredient of recent studies of
dilepton production in heavy-ion collisions due to vector meson dominance. 
It has also been studied in the present model at finite temperature recently
\cite{b+00}.
Here we calculate the $D^*D\pi$ decay process as a quark loop diagram, see Fig.
\ref{ddpi}. 
The decay width of this process is given by 
\begin{eqnarray}
\Gamma_{D^*~ D \pi} = \frac{g^2_{D^*D\pi}}{192 \pi M_{D^*}^5} 
\lambda^3(M_{D^*}^2, M_D^2, M_{\pi}^2)~, 
\end{eqnarray}
with the kinematic factor
$\lambda(s,M_1^2,M_2^2) = \sqrt{[s-(M_1+M_2)^2] [s-(M_1-M_2)^2]}~$.
The $D^*D\pi$ coupling constant $g_{D^*D\pi} = \frac{1}{2}(A - B)$
is given by the functions $A, B$, defined from the transition amplitude of 
this process 
\begin{eqnarray}
T_{D^*~D\pi}^\sigma  &=& A P_\pi^{\sigma} + B P_D^{\sigma} \nonumber \\ 
&=& N_c   \int \frac{d^4p}{(2\pi)^4} 
{\cal F}(p^2)
\mbox{tr}_D \left\{ [ \gamma_\sigma ] 
\left[ -i (p_1 \cdot \gamma) \sigma_V{(p_1^2)} + \sigma_S{(p_1^2)} \right] 
 [i \gamma_5 ] \right.\nonumber \\   
&&\left. \left[ -i ( p_2 \cdot \gamma )  \sigma_V{(p_2^2)}  
+ \sigma_S{(p_2^2)}\right] [i \gamma_5 ]
\left[ -i ( p_3 \cdot \gamma )  \sigma_V{(p_3^2)}  + \sigma_S{(p_3^2)}\right] 
 \right\},
\end{eqnarray}
where ${\cal F}(p^2) = N_{D^*} N_\pi N_D ~\varphi^3(p^2)$ and
$p_1=p-P_\pi/2$, $p_2=p+P_\pi/2$, $p_3=p+P_\pi/2-P_D$.  
Our result for the coupling constant $g_{D^*D\pi} = 10.7$ agrees well with 
the experimental value $g_{D^*D\pi}^{exp} = 10.0 \pm 1.3$ \cite{pdg}.
The next type of diagram to be calculated is the box diagram that 
occurs at fourth order in the meson field expansion.

\section{$J/\psi$ dissociation cross section as a quark exchange process }

The $J/\psi$ dissociation cross section in a hot and dense medium of 
correlated quarks (mesons) has been studied within a nonrelativistic 
potential model \cite{mbq} for quark exchange (string-flip) processes. 
This calculation gives large energy dependent cross sections with a threshold 
enhancement (peak value about $6$ mb) and an exponential tail. 
It is, however, questionable whether the inadequate treatment 
of the pions in these models spoils the results. The pions which are Goldstone 
bosons of the broken chiral symmetry should be described within a chiral 
Lagrangian model. The corresponding reformulation of the charmonium 
dissociation process given previously \cite{mm98} has recently been extended 
and improved \cite{hag99}. These approaches disregard the quark substructure 
effects which give rise to the nonlocality of the 
formfactors and which are expected to become apparent at finite temperatures 
and densities close to the QCD phase transition.

\begin{figure}[t]
\centerline{\psfig{figure=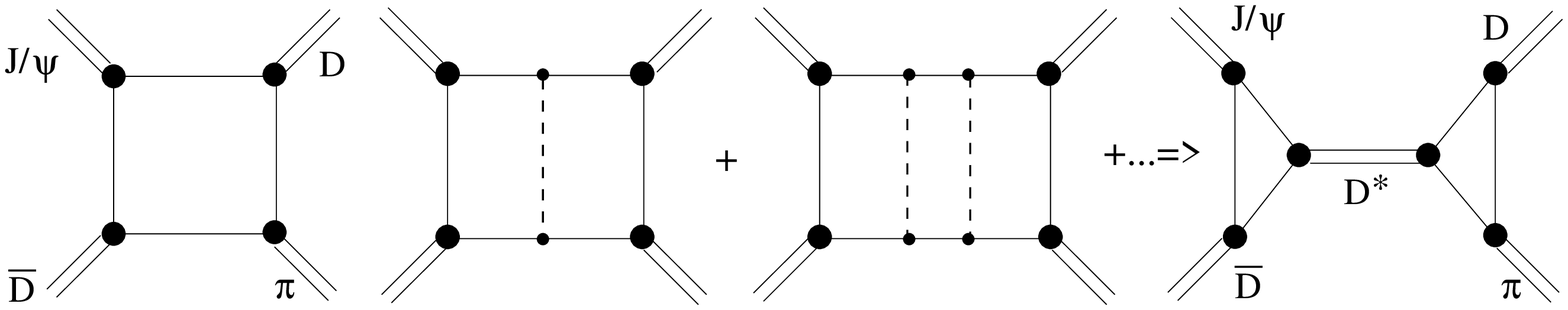,height=3cm,width=12cm}}
\caption{The (anomalous) $J/\psi$ dissociation process in a relativistic quark
model. Quark exchange (box diagram) is found to dominate over D$^*$ exchange 
(ladder resummed box diagram).}
\label{box}
\end{figure}

\begin{figure}[thb]
{\psfig{figure=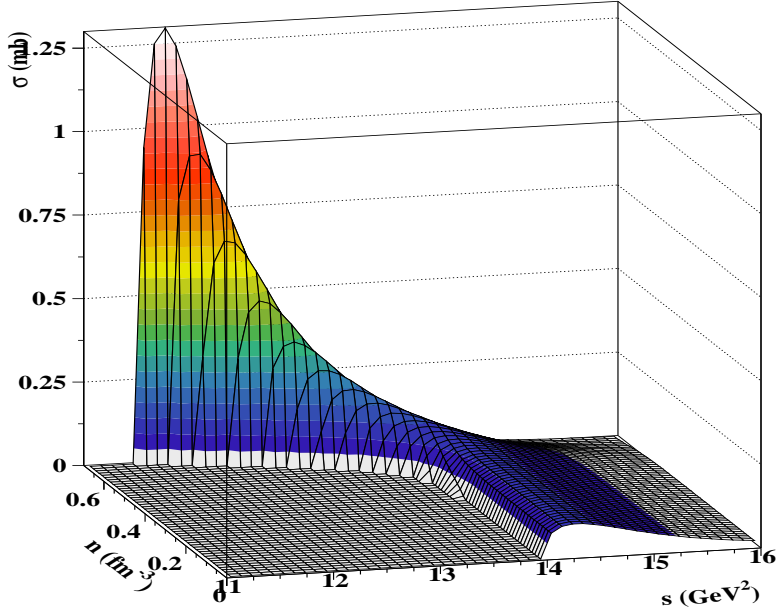,height=10.5cm,width=10.5cm,angle=0}}
\caption{Dependence of the J/$\psi$ dissociation cross section for the 
anomalous process (box diagram) on the squared center-of-mass energy $s$
and on the parton density $n$ of the medium.}
\label{cross}
\end{figure}

Therefore, we consider here the formulation of the $J/\psi$ dissociation 
process within a relativistic quark model at finite temperature. 
As a generic new mechanism on the quark level that we will study in detail is 
the anomalous process 
\mbox{$J/\psi + \pi \to D + {\bar D}$} is described by the
two types of diagram shown in Fig. \ref{box}.
One type represents the quark exchange process 
(box diagram) and the other one describes the $D^*$ exchange (ladder 
resummed box diagram). 
In the language of the effective meson Lagrangian model, the box diagram 
is a contact term which is found to dominate over the heavy-meson exchange.
Therefore we will focus on the evaluation of the cross section for the 
anomalous box diagram as a function of the center of mass energy 
$\sqrt{s}=\sqrt{(P_{{\rm J}/\psi}+P_\pi)^2}$
\begin{eqnarray}
\sigma (s) = \int_{t_{min}}^{t_{max}} dt~ 
\frac{|T_{J/\psi\pi~D \bar{D}}|^2}{16 \pi 
\lambda(s,M_{{\rm J}/\psi}^2,M_\pi^2)}
\end{eqnarray}
with the transition amplitude for this process being 
$T_{J/\psi~\pi D~\bar{D}}$ and the kinematic function $\lambda$ as defined 
above.
In our calculations we put $M_\pi=0$, so that $\lambda=[s-M_{J/\psi}^2],~
s\ge 4 M_D^2~$, and the physical region for this process is determined by
\begin{eqnarray}
t_{\rm min,max} &=& \frac{1}{2}(s - M_{J/\psi}^2)
\biggl[\mp \biggl( 1 - \frac{4 M_D^2}{s}\biggl)^{\frac{1}{2}} - 1\biggl]~
+~ M_D^2~ .
\end{eqnarray}
In the DSE quark model the transition amplitude is given by
\begin{eqnarray}
T_{J/\psi \pi ~ D {\bar D} } &=&  
\epsilon^\mu N_c \int \frac{d^4p}{(2\pi)^4} 
\frac{1}{[p_2^2+m_Q^2][p_3^2+m_Q^2]} 
\nonumber \\ &&
\times \mbox{tr}_D
\biggl\{
\left[ -i ( p_1 \cdot \gamma ) \sigma_V + \sigma_S  \right]
\Gamma_{\bar{D}}
\left[  -i ( p_2 \cdot \gamma )  + m_Q  \right]
\Gamma_{J/\psi}^\mu    
\nonumber \\ 
&&\left[ -i ( p_3 \cdot \gamma )  + m_Q  \right]
 \Gamma_{D}
\left[ -i ( p_4 \cdot \gamma ) \sigma_V + \sigma_S  \right]
\Gamma_\pi   
\biggr\} ~,\label{amplitude}
\end{eqnarray}
where 
$
p_1=p+P_\pi/2, p_2= p+P_\pi/2-P_{\bar{D}}, p_3=p-P_\pi/2+P_{{D}}, 
p_4=p-P_\pi/2
$.
Our result for the s dependence of the cross section 
$\sigma^{\rm box}_{J/\psi \pi~ D {\bar D}}$ is shown in Fig. \ref{cross}. 
We see a critical enhancement just above the reaction threshold. 
Then the cross section drops down for higher values of s. 
This qualitative behavior is similar to the result of a calculation within a 
non-relativistic potential model \cite{mbq} but 
the absolute value of the cross section without in-medium effects ($n=0$)
is much smaller since we have considered the anomalous process first which 
vanishes in the non-relativistic treatment. 

Now we investigate the influence of a hot and dense medium on charmonium 
dissociation. 
A first effect is the modification of the $D$-meson mass $M_D (T)$ due to 
chiral symmetry restoration. 
Similar to the temperature dependence of the light quark mass the 
$D$-mass is nearly constant for lower temperatures and becomes smaller in the 
vicinity of the critical temperature, see Fig. \ref{sigma}.
Secondly, this effect of chiral symmetry restoration enhances 
the cross section for the breakup reaction 
$J/\psi + \pi \longrightarrow D + \bar{D}$. 
The enhancement of the breakup cross section has a critical 
temperature/density dependence at the chiral restoration transition.

\section{J/$\psi$ kinetics at chiral symmetry restoration}

Aiming at a qualitative discussion of possible observable effects of the 
chiral restoration transition to be seen in the J/$\psi$ production cross 
section in ultrarelativistic heavy-ion collisions we consider a relaxation 
time approximation for the evolution of the  J/$\psi$ distribution
in a hot and dense pion gas with the pion distribution function 
$f_{\pi}(\vec{p};T,\mu_\pi)=
\{\exp[(\sqrt{|\vec{p}|^2+M_\pi^2}-\mu_\pi)/T]-1\}^{-1}$. 
This distribution function increases 
strongly near and above the critical temperature.
This leads to a critical enhancement of thermally averaged 
charmonium dissociation cross section
$\left< \sigma_{{\rm J}/\psi \pi~ D \bar{D}} ~v_{\rm rel}\right>$ 
and to a critical drop of the relaxation time $\tau$
\begin{eqnarray}
\tau^{-1}= \left< \sigma_{J/\psi~\pi D \bar D}~v_{\rm rel}\right> n_{\pi}(T) 
= \int \frac{d\vec{p_{\pi}}}{(2\pi)^3} f_{\pi}(\vec{p}_{\pi},T) 
\sigma(\vec{p}_{\pi},\vec{p}_{\psi}) v_{\rm rel} 
\frac{P_{\psi}P_{\pi}}{E_{\psi}E_{\pi}} 
\end{eqnarray}
of the J/$\psi$ distribution in a comoving dense pion (parton) gas.
\begin{figure}[thb]
{\psfig{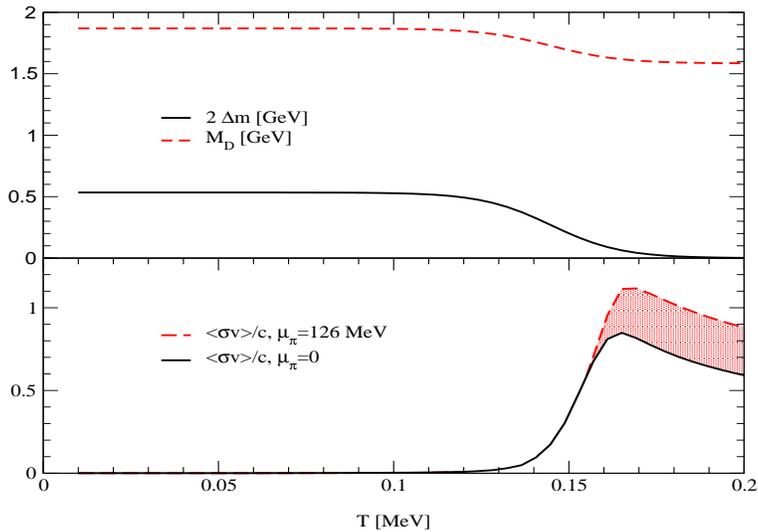}}
\caption{Upper panel: Double quark mass gap (solid line) and D meson mass 
(dashed line) as a function of temperature. 
Lower panel: Temperature dependence of the average 
cross section (arbitrary units) for J/$\psi$ dissociation (anomalous process)
in a hot pion gas with critical enhancement at the chiral restoration 
temperature $T_c=150$ MeV. The dashed band shows the influence of a finite 
(nonequilibrium) pion chemical potential.}
\label{sigma}
\end{figure}
The result is shown in Fig. \ref{sigma} lower panel. 
The critical enhancement of the thermally averaged dissociation cross section 
at the QCD phase transition temperature may lead to the result that in a 
heavy-ion collision above a critical energy density ($E_T$) threshold this 
additional absorption process is switched on and results in an enhanced 
J/$\psi$ suppression. This pattern may serve as an explanation for the 
puzzling observation of anomalous J/$\psi$ suppression in the 
NA50 experiment at CERN SpS, see Fig. \ref{na50}.

\section{Conclusions}
A recently developed dynamical approach to the chiral symmetry restoration and
deconfinement transition is generalized for heavy mesons and 
quarkonia; in-medium modifications of masses and decay constants are 
studied. 
We apply this model to calculate the cross section for charmonium dissociation
by hadron impact and find that contributions from the quark 
exchange  dominate over those from D$^*$ resonance propagation processes.
At the QCD phase transition the reaction rates for the J/$\psi$ breakup 
are strongly enhanced.

The approach will be further developed in order to include 
the $\rho$-meson which catalyzes exothermic J/$\psi$ breakup and 
a treatment of the normal process 
J/$\psi + \pi \to D \bar{D^*}, D^*\bar{D}$, 
which should be larger than the anomalous one considered here.
On this basis, a detailed quantitative calculation of  J/$\psi$ production in 
the NA50 experiment can be attacked and predictions for the RHIC and LHC 
colliders can be made. A particularly new aspect is the occurrence of 
$D \bar{D}$ annihilation into $J/\psi + \pi (\rho)$ (gain process) 
under these conditions which modifies the charmonium kinetics \cite{ko+98,pbm}
and may lead to a saturation or even enhancement of the J/$\psi$ abundance.

\section*{Acknowledgments}
The authors thank P. Braun-Munzinger, J. H\"ufner, B. M\"uller, C.D. Roberts,
G. R\"opke and S.M. Schmidt for stimulating discussions.  
MII and YLK acknowledge the hospitality of the Physics Department at the 
University of Rostock where part of this work has been done; GB was
supported by a stipend from the Max-Planck-Gesellschaft. 
This work has been supported by the Heisenberg-Landau 
programme, the Deutscher Akademischer Austauschdienst (DAAD), 
the Russian Fund for Fundamental Research, contract number 97-01-01040,
the National Science Foundation under Grant No. INT-9603385 and by the 
DFG-Graduiertenkolleg ``Stark korrelierte Vielteilchensysteme'' at the 
University of Rostock. 

\end{document}